# Towards quantification of the ratio of the single and double wall carbon nanotubes in their mixtures:An In situ Raman spectroelectrochemical study

Zuzana Kominkova[a,b], Vaclav Vales[a], Mark C. Hersam[c], Martin Kalbac [a,*]

[a] J. Heyrovský Institute of Physical Chemistry, Academy of Sciences of the Czech Republic, Dolejškova 3, CZ-18223 Prague 8, Czech Republic

[b] Palacký University, Department of Physical Chemistry, Faculty of Science, Tř.17.listopadu 12, CZ-77146 Olomouc, Czech Republic

[c] Department of Materials Science and Engineering and Department of Chemistry, Northwestern University, Evanston, Illinois, 60208-3108, United States.

**Abstract**

Mixtures containing different weight ratios of single wall carbon nanotubes (SWCNT) and double wall carbon nanotubes (DWCNT) were prepared and studied by in-situ Raman spectroelectrochemistry. Two components of the G' mode in the Raman spectra, which can be resolved at high electrode potentials, were assigned to the signals from inner tubes of DWCNT and outer tubes of DWCNT together with SWCNT. The dependence of the ratios of these two components of the G' mode on the nominal amount of SWCNT and DWCNT in the samples was simulated so that the residual amount of SWCNT in the original DWCNT could be determined. Additionally, the individual contributions of all components of carbon nanotubes into the total area of the G' mode at high electrode potentials were estimated from the simulation.

## 1. Introduction

Double wall carbon nanotubes (DWCNT) are a special case of multi walled carbon nanotubes. The outer wall protects the inner one from the environment, which allows the inner tube to keep its unique properties while the outer tube can be modified to improve the adhesion of nanotubes to polymers in composites [1], to facilitate targeting of nanotubes for drug delivery [2,3], or to enhance selectivity and sensitivity of nanotube-based chemical sensors [4]. This growing list of DWCNT applications has motivated the development of methods to grow, separate, and purify DWCNT [5,6].

To accelerate improvements in the growth, separation, and purification of DWCNT, methods for quantifying the purity of DWCNT are needed. Since the main contaminants in DWCNT samples are usually single wall carbon nanotubes (SWCNT), this task is challenging due to difficulties in distinguishing between individual types of CNT. High resolution transmission electron microscopy (HRTEM) is practically the only technique that allows quantification of the amount of DWCNT compared to SWCNT by direct counting [7,8]. However, this procedure is time consuming and thus only a small fraction of the sample can be probed by HRTEM.

* Corresponding author.
Tel.: +420 2 6605 3804. E-mail: martin.kalbac@jh-inst.cas.cz (M. Kalbac).

Raman spectroscopy is a sensitive characterization method for carbon-based materials, such as carbon nanotubes, graphene and peapods, due to the strong resonant enhancement of the Raman signal [9]. It provides information about diameter distribution [10], electronic type of nanotubes [11], defects [12–14], strain [15,16] and/or doping [17,18]. The important Raman active bands, which appear in the Raman spectra of carbon nanotubes, are the radial breathing mode (RBM), the tangential displacement mode (TG), the disorder induced mode (D) and the high frequency, two phonon mode (G').

The radial breathing mode is observed as a peak in the 100-300 cm$^{-1}$ range and is often considered as a fingerprint for SWCNT because its frequency ($\omega_{RBM}$) is related to the tube diameter $d$ [19],:

$$\omega_{RBM} [cm^{-1}] = 217.8/d [nm] + 15.7 \text{ cm}^{-1} \tag{1}$$

The RBM mode is observed both for inner and outer tubes of DWCNT, however it is difficult to distinguish between the inner tubes of DWCNT and thin SWCNT.

Two other prominent features of carbon nanotubes, the tangential (TG) and the high frequency mode (G'), are only weakly dependent on the nanotube diameter [20], hence the position of these modes provides little to no information about the diameter distribution of the tubes.

It has been shown previously that combining electrochemical doping and Raman measurements of carbon nanotubes can provide significantly more insight into the understanding of carbon nanotube properties [21–23]. In contrast to chemical doping, electrochemistry allows reproducible, precise and well-controlled doping of carbon nanotubes. In particular, electrochemical charging leads to a shift of the Fermi level. As the energy of a van Hove singularity is reached, the corresponding interband transition is bleached and the Raman resonance is quenched [17]. In contrast to single wall carbon nanotubes, the electrochemical doping of double wall carbon nanotubes is more complex [24,25]. For example, it has been shown that the effects of doping of inner tubes depend on specific combinations of electronic structure of the outer and inner tubes. These effects are well distinguishable in samples sorted by electronic type [26]. Nevertheless, even for samples with DWCNT of mixed electronic types, one may follow the general trend that the effect of electronic charge is stronger for outer tubes than inner tubes [24,25].

Here, we focus on Raman spectroelectrochemical measurements of SWCNT, DWCNT and their mixtures. The subsequent analysis of the data allowed us to construct a calibration curve, i.e. the dependence of the relative integral intensity of the upshifted peak on the relative content of SWCNT in the mixture. The calibration curve thus serves as a tool for determining the proportion of SWCNT compared to DWCNT in a mixed sample.

1. **Experimental**

2.1 *Sample preparation and characterization*
The samples of SWCNT and DWCNT were obtained from Unydim, USA and Thomas-Swan, UK. For the analysis, mixtures of SWCNT and DWCNT (4:1, 2:1, 1:1, 1:2 and 1:4 SWCNT to DWCNT ratios) were prepared by adding SWCNT into the commercial DWCNT sample. The electrochemical measurements were performed using a three electrode system connected to an Autolab PGSTAT (Ecochemie). In order to obtain homogeneously dispersed tubes, the probed CNT mixtures were sonicated in ethanol. A drop of such dispersion was put on a Pt wire which after evaporation of ethanol served as a working electrode. The counter electrode was represented by another Pt-wire and Ag-wire was employed as a reference electrode. The electrolyte solution was $LiClO_4$ (0.2M) in dry acetonitrile (both from Aldrich). The in-situ Raman spectroelectrochemistry of the SWCNT/DWCNT mixture in different quantity ratios was investigated in the electrode potential range between -1.5 V and 1.5 V (in steps of 0.3 V) vs. Ag and between 1.0 V and 1.5 V (in steps of 0.1 V) vs. Ag. Measured CNT sample should be thin due to better effect of doping by electrode potential. The Raman spectra were excited by a 1.96 eV laser, and the spectra were recorded by a Labram HR spectrometer (Horiba Jobin Yvon) interfaced to an Olympus microscope (objective 50x). The laser power impinging on the cell window was about 1 mW.

2.2 *Description of the simulation model*
For the analysis of the measured Raman data of the SWCNT/DWCNT mixtures measured at different potentials, we focused on the evolution of the G' mode. We assumed that SWCNT and the outer tubes of DWCNT (o-DWCNT) are affected by the electric potential in the same way and therefore their G' modes are shifted equally. On the other hand, the inner tubes of DWCNT (i-DWCNT) are nominally protected from the electric field by the outer tubes, which implies that its G' mode is not shifted and can be distinguished from the shifted G' mode. The relative area of the G' mode corresponding to SWCNT and o-DWCNT normalized to the total area of G' mode (*A*) thus depends on the relative amount of SWCNT in the mixture (*m*). We assume the following model to simulate the dependence of *A* on *m*,

$$A_{sim} = \frac{[m + (1-m)t]w_s + (1-m)(1-t)w_o}{[m + (1-m)t]w_s + (1-m)(1-t)(w_o + w_i)}, \quad (2)$$

where *t* is the relative amount of SWCNT in nominally pure DWCNT, and the parameters $w_s$, $w_o$, $w_i$ describe the contributions of individual tubes (SWCNT, o-DWCNT, and i-DWCNT) to the G' mode. While $w_s$, $w_o$, $w_i$ are assumed to be constant for different mixtures at a single potential, *t* is supposed to be constant for both all mixtures and potentials. All of these parameters can be determined from the fit of this model to the measured data.

## 2. Results and discussion

In our study, we probed SWCNT, DWCNT and their mixtures by in-situ Raman spectroelectrochemistry. Figure 1 shows Raman spectra of the pure SWCNT sample obtained at different electrode potentials. The frequencies of the RBM bands of the SWCNT sample are 197, 218, 258 and 283 cm$^{-1}$, and according to equation (1), the RBM bands correspond to tube diameters of about 1.20, 1.08, 0.90 and 0.82 nm, respectively. Electrochemical charging leads to an overall bleaching of the intensity of the Raman signal. The bleaching is roughly symmetric for both cathodic and anodic doping, which reflects the mirror pairs of van Hove singularities. The electrochemical charging of the carbon nanotubes causes a shift of the Fermi level. According to a simple model, as soon as the Fermi level reaches the level of a van Hove singularity it erases the electronic transitions from/to this particular singularity. If the Raman process is in resonance with such a singularity, strong bleaching of the Raman signal is expected. Previous detailed studies showed that the bleaching occurs already when any electronic state of SWCNT is filled/depleted [17,27]. This effect can easily be seen in the case of the RBM mode since it has a relatively narrow resonance window [28].

The frequency of the TG mode is found at 1593 cm$^{-1}$ in SWCNT without any electrode potential. During positive charging, the TG mode of semiconducting tubes shows an upshift which depends on the tube diameter because of the simultaneous action of the changed force constant and the phonon renormalization effect [29]. The TG mode is shifted in frequency (to 1604 cm$^{-1}$) and broadened at higher positive potentials (1.0-1.5 V). The attenuation of the Raman spectra (excited at 1.96 eV) is usually explained by the loss of resonance through quenching of the optical transitions $E^M_{11}$ and $E^S_{22}$ in narrower tubes.

At higher positive potentials (1.0-1.5 V) a broadening of the G' mode has also been observed. Smaller splitting of the G' mode was previously observed also for isolated metallic SWCNT, indicating a gentle splitting of the van Hove singularities due to the trigonal warping effect [30,31]. This effect is quenched in nanotube bundles and can therefore be excluded in our case.

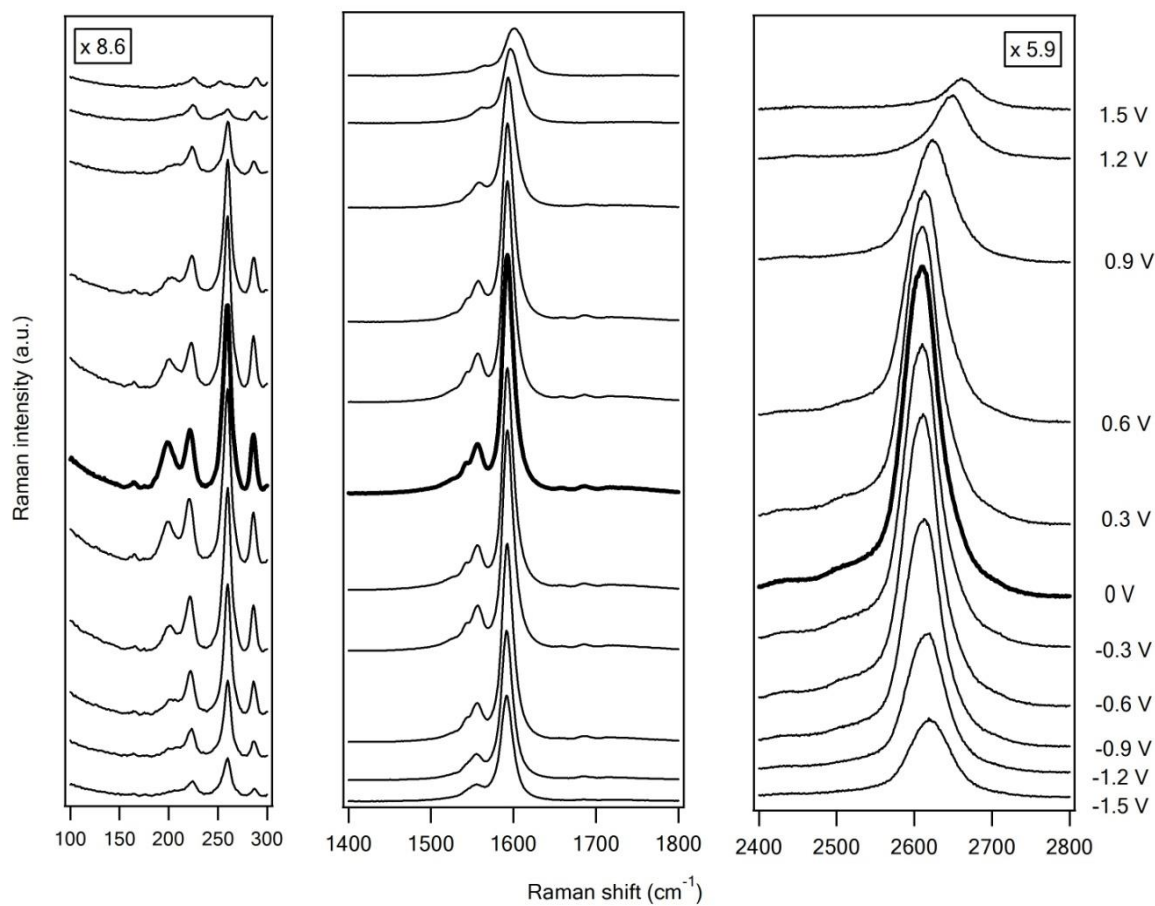

**Figure 1.** In-situ Raman spectroelectrochemical data of SWCNT (HiPco) in the electrode potential range from -1.5 to 1.5 V. The spectra are excited by 1.96 eV (633 nm) laser radiation. The electrochemical potential is labeled next to each curve.

The Raman bands in electrochemically charged DWCNT are shown in figure 2. It is important to note that this DWCNT sample also contains a portion of SWCNT. According to equation (1), the RBM bands with frequency at 153, 197, 215, 253 and 283 $cm^{-1}$ correspond to tube diameters of about 1.59, 1.20, 1.10, 0.92 and 0.82 nm, respectively. The response of DWCNT to the electrode potential is different as compared to SWCNT, since the bleaching of the RBM bands depends on whether the particular tube is inner or outer [32]. The fast bleaching mode of the RBM bands at 153 $cm^{-1}$ was assigned to outer tubes, the peaks at 197 and 283 $cm^{-1}$ were assigned to SWCNT while the slow bleaching modes at 215 and 253 $cm^{-1}$ were tentatively assigned to the inner tubes. The slow bleaching of the Raman signal of the inner tubes is in agreement with previous studies [24–26,32] and it can be rationalized by a delayed charge transfer from outer to inner tube [26]. A previous study [29] of SWCNT showed an upshift of the TG mode during positive charging due to a change of the force constant and the phonon renormalization effect. However, the latter effect is postponed for inner tubes. Therefore the TG bands of the inner and outer tubes of DWCNT can be distinguished at high anodic potentials, which results in a characteristic doublet of the peak of the TG mode. A broadening and splitting of the G' mode has also been observed at higher positive potentials.

The doping-induced splitting of the TG and G' modes is specific for DWCNT and can therefore be used for evaluation of the purity of DWCNT. The DWCNT in our sample have diameters of outer and inner tubes in the range of about 1.6 and 0.8–0.9 nm, respectively. These two regions of diameters manifest themselves as two distinct G' modes. Hence, the low frequency component corresponds to inner tubes and the high frequency component to outer tubes. The inner and outer tubes correspond to different electronic transitions, which are involved in their resonance process. The outer tubes are resonant enhanced through the $E_{33}$ transition, while the inner tubes through the $E_{22}$ transition at an excitation energy of 1.96 eV. Therefore, the TG and G' modes splitting at positive potentials can be used as another indication of DWCNT in the sample, as is shown in figure 2.

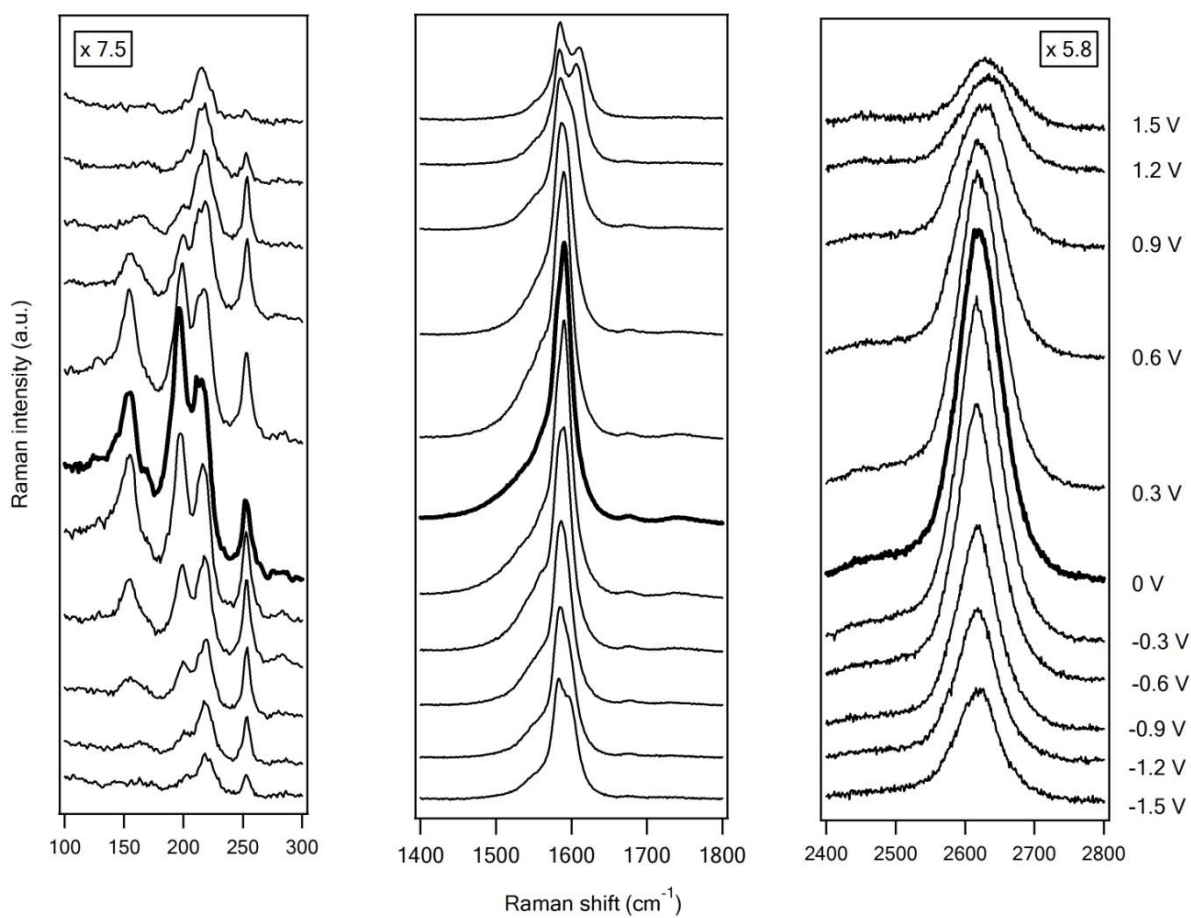

**Figure 2.** In-situ Raman spectroelectrochemical data of a DWCNT sample in the electrode potential range from -1.5 to 1.5 V. The spectra are excited by 1.96 eV (633 nm) laser radiation. The electrochemical potential is labeled next to each curve.

In order to evaluate the amount of SWCNT vs. DWCNT one would suggest using the RBM, TG and G' since potential induced changes are observed for all these bands. The RBM intensity dependence on potential is however significantly different for tubes on and off resonance due to the relatively narrow resonance window for the RBM band [28]. In addition the diameter of the tubes plays a significant role in assignment of RBM bands to SWCNT and DWCNT. Generally, the bleaching of tubes with larger diameter is faster since the van Hove singularities are closer to the Fermi level. In our previous study[33] we have shown that in situ Raman spectroelectrochemical study can differentiate SWCNT from the inner tubes of DWCNT. The intensity of RBM band of inner tubes of DWCNT remains almost unchanged upon the electrochemical charging while the RBM intensity of SWCNT is bleached fast. Furthermore, at the high anodic potential the splitting of the TG mode of DWCNT is observable, therefore we are able to resolve DWCNT from SWCNT. In this study we employ qualitatively known different behavior of SWCNT and DWCNT to quantify the relative amount of both types of tubes. On the other hand, both the TG and G' modes exhibit relatively broad resonance windows and both split at high anodic potentials, hence the inner and outer tube can be resolved. It was previously found that the combination of electronic type of the inner and outer tubes in DWCNT strongly affects the charge induced changes of the TG mode of the inner tubes [26]. This would again complicate the quantification of the SWCNT/DWCNT ratio. On the other hand, the G' mode dependence on electrode potential has been found to be beneficial for the distinguishing outer tube and inner tube of DWCNT [34]. In addition, the G' mode shift is relatively strong at high positive potentials (see figure 1). Therefore, the G' mode is a good candidate for the quantification of the ratio of the SWCNT/DWCNT ratio. A study of the evolution of the G' mode as a function of different SWCNT/DWCNT ratios under chemical doping has been already published [35]. However, the model presented there does not consider possible contamination of initial DWCNT with SWCNT which often occurs. To cope with this issue an electrochemical doping with several, well defined potential applied to the sample, turned out to be necessary.

In order to evaluate the SWCNT/DWCNT ratio, we prepared mixtures of SWCNT and DWCNT in ratios of 4:1, 2:1, 1:1, 1:2 and 1:4 and each of them was analyzed by in-situ Raman spectroelectrochemistry. Each measurement was repeated five times at different locations on the sample and the data were averaged in order to get better statistics. The G' mode broadening increases with increasing anodic potential. On the other hand, the intensity decreases with increasing electrode potential, which in turn complicates the decomposition of the G' mode to the contribution of SWCNT and DWCNT. For this reason, we focused on the potential range between 1.0 and 1.5 V since the peak splitting is observable and the decrease of the intensity of the G' mode is reasonable. The G' mode was fitted by two peaks with a Lorentzian line shape, where the low frequency peak was assigned to the inner tube of DWCNT and the high frequency peak to the outer tubes of DWCNT and to SWCNT.

In order to obtain reliable results, several fitting constraints were applied. Since the mixing of two types of carbon nanotubes should not affect their individual properties, the only parameter that is expected to change for different mixing ratios at any given potential is the ratio of the integral intensity of both peaks. Therefore, the full width at half maximum (FWHM) of the low-frequency and the high-frequency peaks and the spectral shift between these two peaks were kept the same for all the mixtures at each potential.

An example of the fits of the G' mode with two peaks at 1.5 V electrode potential is shown in figure 3. The G' mode is partly overlapping with the G* mode (at around 2450

cm$^{-1}$). For this reason, the G* mode had to be fitted as well, however it was not used for the analysis. Figure 3 shows the peak evolution for both bands of the G' mode with increasing amount of SWCNT in the mixtures at 1.5 V electrode potential. With increasing amount of SWCNT in the mixture, the relative intensity of the high frequency peak increases and the intensity of the low frequency peak, which corresponds to the inner tube of the DWCNT, decreases. The position of the low frequency component of the G' mode is at about 2605 cm$^{-1}$. The high frequency component of the G' mode of the DWCNT is blue shifted by about 37 cm$^{-1}$ to 2642 cm$^{-1}$. The frequency of the G' mode of the inner tubes of the DWCNT does not change with increasing electrode potential, since they are shielded by the outer tubes.

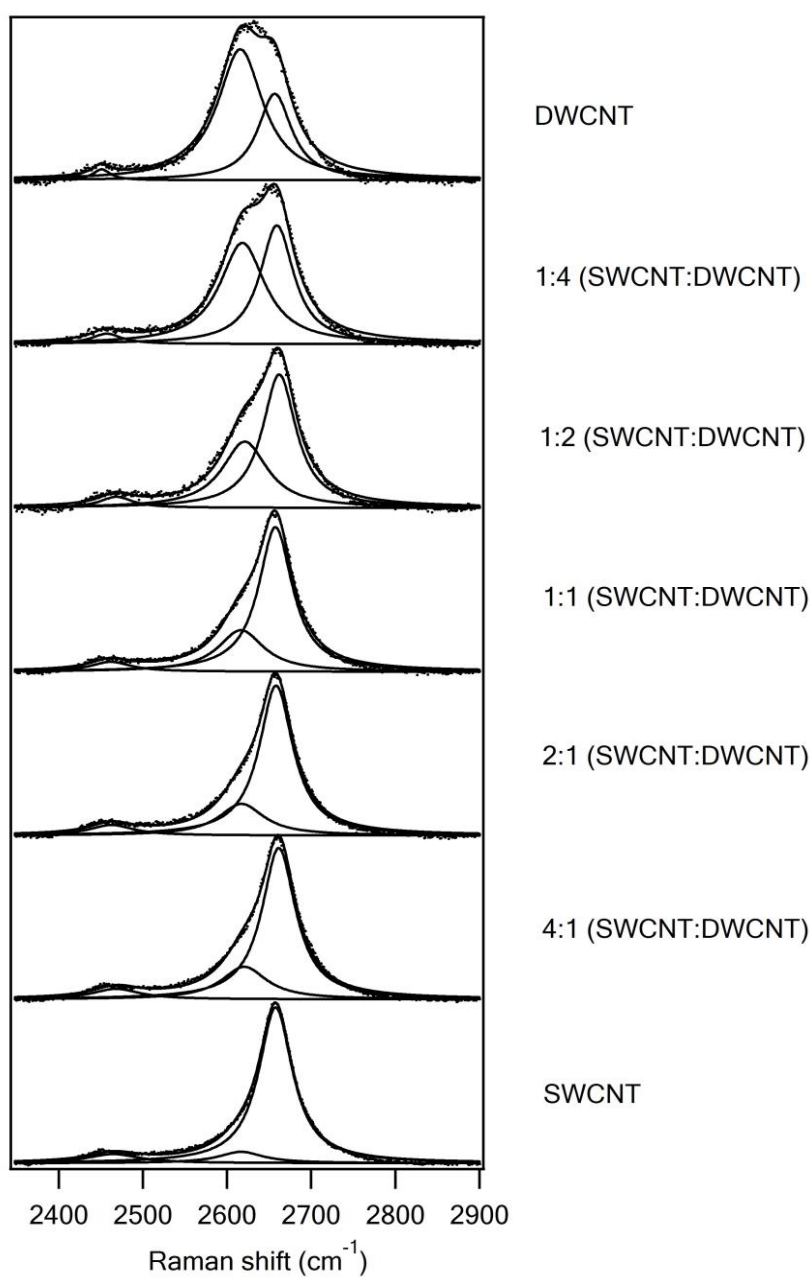

**Figure 3.** The G' mode and their fits for different ratios of DWCNT to SWCNT in samples measured at 1.5 V electrode potential. The dotted lines represent the experimental points and the solid lines their fit with a Lorentzian line shape.

Figure 4 shows the Raman shift between the low and the high frequency components of the G' mode for each measured value of electrode potential fitted for all mixtures together. As we already mentioned, the broadening of the G' mode increases with increasing positive potential. This effect is in accordance with our results. With increasing anodic potential, the Raman shift between the two bands of the G' mode is increasing.

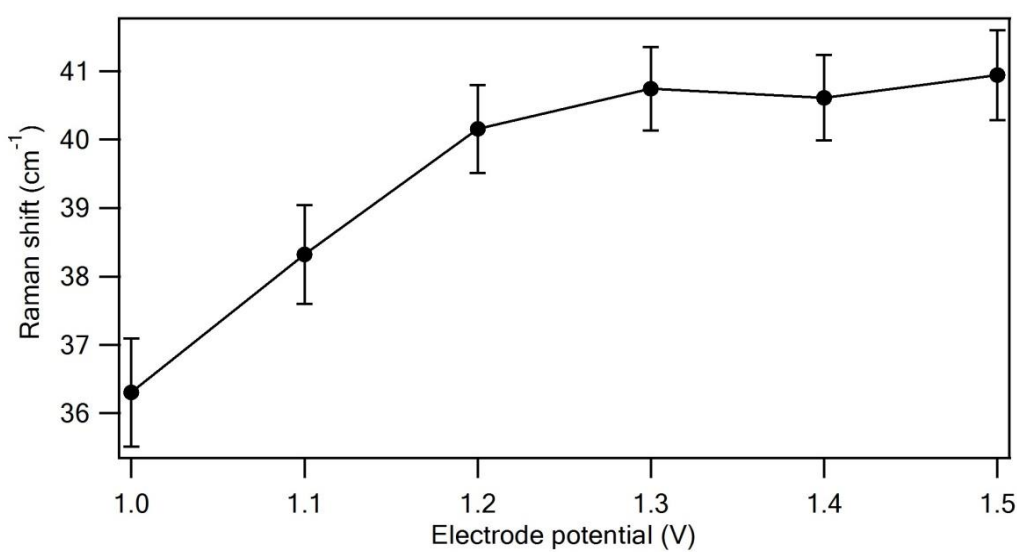

**Figure 4.** Raman shift between the two bands of the G' mode at different electrode potentials (1.0-1.5 V). The error bars from the fits are depicted.

For an improved characterization of the behavior of the two bands of the G' mode at applied electrode potential, we also focused on the FWHM of each band at different positive potentials fitted for all mixtures together (see figure 5). From the curve corresponding to the low frequency peak, it is evident that its FWHM increases to 1.4 V and is then slightly decreased, but this narrowing is within the fitting errors. This effect is the opposite in the case of the FWHM of the high frequency peak. With increasing electrode potentials, the FWHM of the high frequency peak decreases to 1.3 V and then increases. These changes at the highest electrode potential could be caused by the weak Raman signal at 1.5 V.

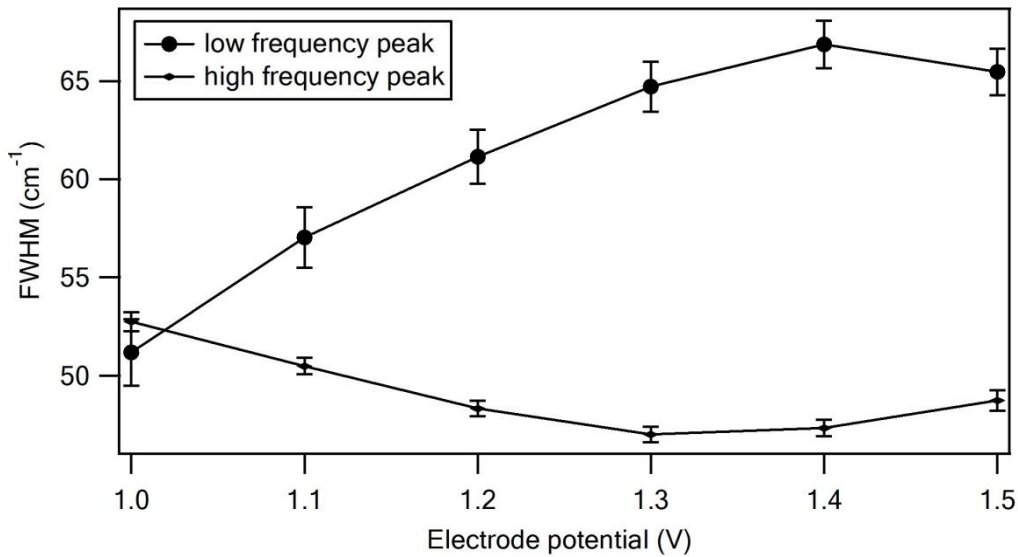

**Figure 5.** Evolution of the full width at half maximum (FWHM) of the two bands of the G' mode with increasing electrode potential. The error bars from the fits are depicted.

In the next part of our study, we calculated the content of SWCNT in the original DWCNT (see figure 6). We fitted the dependence of the ratios of the high frequency peak to the total area of both bands of the G' mode (*A*) on nominal ratios of SWCNT vs. total amount of tubes (*m*) for each mixture of SWCNT vs. DWCNT using the simulation described above. In the simulation the relative amount of SWNCTs in nominally pure DWCNT (*t*) and three parameters corresponding to the contributions of each individual carbon nanotube type, i.e. SWCNT, i-DWCNT and o-DWCNT ($w_s$, $w_i$, $w_o$), were taken into account. As already mentioned, at the highest electrode potentials the G' mode is more broadened and decreased in intensity in comparison with the G' mode at lower electrode potentials. This effect complicates the decomposition of the G' mode to the contribution of SWCNT and DWCNT.

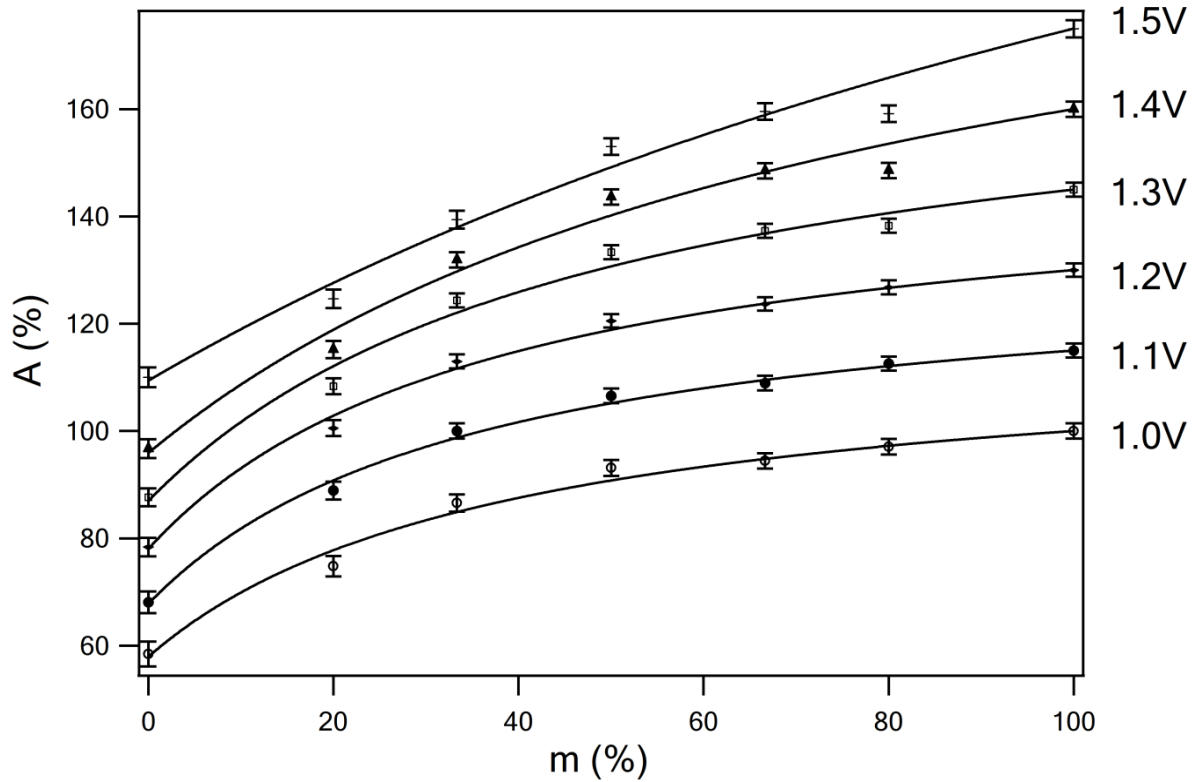

**Figure 6.** The dependence of the ratios of the G' peak corresponding to SWCNT and outer tubes of DWCNT normalized by the total area of the G' mode peaks (A) on different nominal ratios of SWCNT vs. total amount of nanotubes in the samples (m) in the range of electrode potentials between 1.0 and 1.5 V (in steps of 0.1 V). The curves were upshifted for better clarity. The error bars from the fits are depicted.

From the data obtained from fits of the two bands of the G' mode at different electrode potentials, we were able to calculate the contributions of each carbon nanotube type into the G' mode. As mentioned above, we assumed that the low frequency peak corresponds to the inner tubes of DWCNT and the high frequency peak corresponds to the outer tubes of DWCNT together with SWCNT. Since the inner tubes should be unaffected by the electric potential, the contributions to the G' intensity of the outer tubes of the DWCNT and the SWCNT, respectively, were normalized to the intensity of the inner tubes of the DWCNT. The dependences of the individual area of the contributions of carbon nanotubes are plotted in figure 7. From figure 7 it is obvious that the SWCNT contribute significantly more to the area of the G' mode than the outer tubes of DWCNT. These results are in accordance with the Raman spectra in figure 1 and 2. From the RBM modes of the SWCNT and the DWCNT, it can be seen that the outer tubes of the DWCNT have the largest diameter and in these tubes, the van Hove singularities are closer to each other than in the case of carbon nanotubes with smaller diameters. For this reason, the outer tubes of the DWCNT are bleached faster than the other carbon nanotubes with smaller diameters in our sample and at high electrode potentials their contributions are almost negligible. From the plots corresponding to the SWCNT, and also to the DWCNT to a lesser extent, a drop of the contributions into the G' mode with increasing positive electrode potential is evident. The amount of SWCNT in initial DWCNT has been determined from the fit to $t = 10\ \%$. However, the parameter $w_o$ that describes the contribution of the outer tubes of the DWCNT, strongly correlates with the amount of the SWCNT in initial nominally pure DWCNT. Nevertheless it can be said that there is a maximum 14 % of SWCNT in nominally pure DWCNT. The producer approximated 15 % of SWCNT in DWCNT, which is in a good match to our data. The sensitivity of the purity evaluation depends on the quality of the measured data. In our case we can say from the comparison of the result obtained by our method and from TEM that the detectable amount of SWCNT in SWCNT/DWCNT mixture is around 5%.

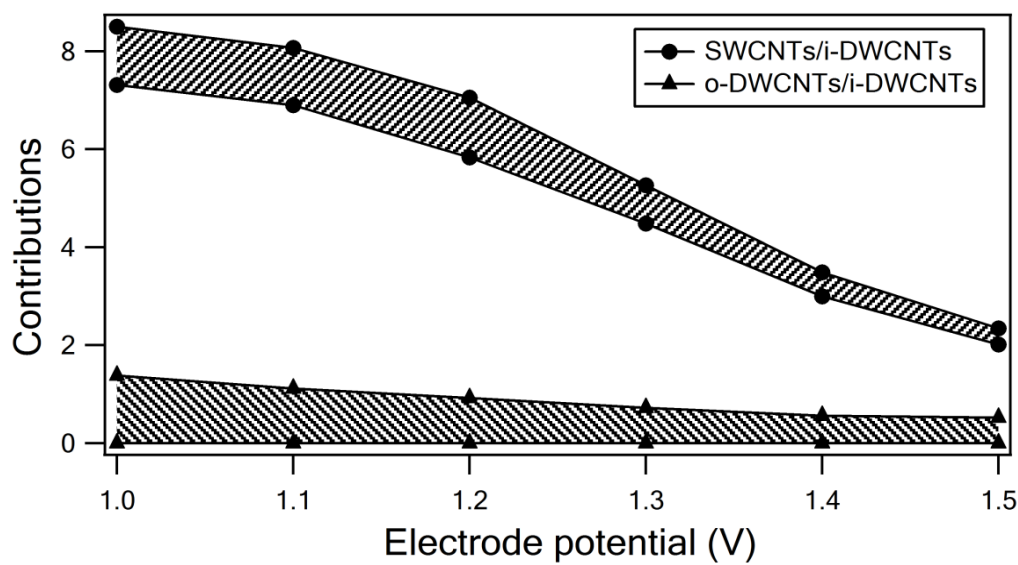

**Figure 7.** Area ratio of the G' band of outer tubes of DWCNT (o-DWCNT) and SWCNT relative to inner tubes of DWCNT (i-DWCNT) at different electrode potentials. In the graph the upper and lower limits are drawn for individual contributions.

3. **Conclusions**

We developed a new procedure to evaluate the relative amount of DWCNT and SWCNT in their mixtures. The method is based on analysis of the G' mode in the Raman spectra in electrode potentials from 1.0 to 1.5 V. The G' mode has been found to split with increased electrode potential due to a different dependence of the G' mode frequency of inner tubes of DWCNT and SWCNT on electrode potential. From the experimental ratio of the low and high frequency peaks of the G' mode, a calibration curve was constructed and included to evaluate the amount of DWCNT in the original sample. The average proportion of SWCNT in the original DWCNT was up to 14%. We also focused on the evolution of the individual contributions of each carbon nanotube type into the total area of the G' mode at high electrode potentials.

**Acknowledgements**

The authors thank to Dr. M. Motta for providing DWCNT/SWCNT samples. The work was supported by the MSMT (contract No., LL1301) and Czech Grant agency (P204/10/1677). M.C.H. acknowledges funding from the National Science Foundation (DMR-1006391).